# Incomplete beta-function expansions
# of the solutions to the confluent Heun equation

**Artur Ishkhanyan**

Engineering Center of Armenian National Academy of Sciences,
Ashtarak-2, 378410 Armenia

**Abstract**

Several expansions of the solutions to the confluent Heun equation in terms of incomplete Beta functions are constructed. A new type of expansion involving certain ***combinations*** of the incomplete Beta functions as expansion functions is introduced. The necessary and sufficient conditions when the derived expansions are terminated, thus generating closed-form solutions, are discussed. It is shown that termination of a Beta-function series solution always leads to a solution that is necessarily an elementary function.

The Heun equation (a second order Fuchsian linear differential equation with four regular singular points) [1] generates by confluence four different differential equations with irregular singularities. The singly confluent Heun equation is the first confluent equation obtained from the general Heun equation when the singularity at a finite point $z=a$ of the complex $z$-plane is merged with that at infinity, thus resulting in an irregular singularity of rank $1$ at infinity. The canonical form of this equation is written as [2,3]

$$u'' + \left(4p + \frac{\gamma}{z} + \frac{\delta}{z-1}\right)u' + \frac{4p\alpha z - \sigma}{z(z-1)}u = 0. \tag{1}$$

This equation is widely known both in physics and mathematics, and it has been extensively investigated from different points of view. Svartholm and Erdélyi initiated a powerful technique of solution of the Heun equation by means of series of hypergeometric functions [4,5]. This technique has been applied to the confluent Heun equation (1) by numerous authors. Expansions in terms of Gauss hypergeometric and confluent hypergeometric functions have been constructed (see, e.g., [6,7]) and applied to different problems (for a survey, see [2]).

In the present letter, we show that when $\alpha = 0$ or $\sigma = 0$ expansions in terms of incomplete Beta functions [8] are also applicable. Furthermore, we show that in some cases new expansions can be constructed using as expansion functions certain ***combinations*** of Beta functions (such expansion functions has been introduced in our earlier papers [9] to construct solutions to the general Heun equation). In particular, we develop such an expansion for the case $\sigma = 4p\alpha$. The mentioned expansions are governed by three-term recurrence relations between successive coefficients of the expansion. We discuss the conditions for termination of these expansions and write out the first two explicit closed-form solutions derived for each case of termination. Further, we construct a generalized expansion involving combinations of Beta functions for the case $\alpha = 0$. The expansion turns out to obey a four-term recurrence relation. Being different in general, however, this expansion generates the same closed form solutions as the above first expansion for $\alpha = 0$ governed by a three-term recurrence relation. Finally, we show that several more elaborate developments resulting in different Beta-function expansions for other (non-trivial) values of involved parameters are also possible.

Using an alternative representation of the incomplete Beta function through the Gauss hypergeometric function, it can easily be shown that the presented finite-sum solutions are always expressed in terms of elementary functions. We discuss this question in general and show that any solutions in the form of a linear combination of a finite number of incomplete Beta functions should necessarily be elementary functions. It should be mentioned here that there have been several recent studies where incomplete Beta functions appeared in solving the Heun equations (see, e.g., [10]). We would like to note that the solutions obtained are particular cases of the finite-sum expansions considered here and that in fact, according to the aforesaid, these are elementary-function solutions.

We search for expansions of the solutions to the confluent Heun equation in the form

$$u = \sum_n a_n u_n, \quad u_n = B_z(\gamma_0 + n, \delta_0), \qquad (2)$$

where $B_z(\gamma_n, \delta_0)$ (hereafter $\gamma_n = \gamma_0 + n$) is the incomplete Beta function that is defined as [8]

$$B_z(a,b) = \int_0^z t^{a-1}(1-t)^{b-1} dt, \quad \mathrm{Re}(a) > 0. \qquad (3)$$

This function is written in terms of elementary functions when $a$ (or $b$) is positive integer or when $a+b$ is a negative integer. The latter property is easily seen from the following alternative representation of the incomplete Beta function through the Gauss hypergeometric function [8]:

$$B_z(a,b) = \frac{z^a(1-z)^b}{a} {}_2F_1(1, a+b, 1+a, z). \qquad (4)$$

The functions $u_n$ obey the differential equation [$\mathrm{Re}(\gamma_n) > 0$]

$$u_n'' + \left(\frac{1-\gamma_n}{z} + \frac{1-\delta_0}{z-1}\right)u_n' = 0. \qquad (5)$$

Substitution of (2) and (5) into (1) gives

$$\sum_n a_n \left[ \left( 4p + \frac{\gamma - 1 + \gamma_n}{z} + \frac{\delta - 1 + \delta_0}{z - 1} \right) u'_n + \frac{4\alpha p z - \sigma}{z(z-1)} u_n \right] = 0 \qquad (6)$$

or

$$\sum_n a_n \left[ \left( 4p z(z-1) + (\gamma - 1 + \gamma_n)(z-1) + (\delta - 1 + \delta_0) z \right) u'_n + (4\alpha p z - \sigma) u_n \right] = 0. \qquad (7)$$

Between the involved incomplete Beta functions the following recurrence relations hold

$$z(z-1) u'_n = -\gamma_n u_n + (\gamma_n + \delta_0) u_{n+1}. \qquad (8)$$

$$(z-1) u'_n = -\gamma_{n-1} u_{n-1} + (\gamma_{n-1} + \delta_0) u_n. \qquad (9)$$

Hence, the first two terms in equation (7) are always expressed in terms of functions $u_n$. However, a key observation for what follows is that the combination $z(C_1 u'_n + C_2 u_n)$ is not expressed as a linear combination of functions $u_n$ for any (nonzero) $C_{1,2}$. There are different possibilities to overcome this difficulty. Below we present several such developments.

**1.** A straightforward possibility is to equate the term $z((\delta - 1 + \delta_0) u'_n + 4\alpha p u_n)$ to zero thus supposing that parameter $\alpha$ of the initial Heun equation is zero (obviously, the alternative case $p = 0$ is a trivial one since then the Heun equation is simply reduced to the Gauss hypergeometric equation) and further putting $\delta_0 = 1 - \delta$. We then get a three-term recurrence relation between the successive coefficients of expansion (2):

$$R_n a_n + Q_n a_{n-1} + P_n a_{n-2} = 0, \qquad (10)$$

where

$$R_n = (\gamma - 1 + \gamma_n) \gamma_{n-1}, \qquad (11)$$

$$Q_n = 4p \gamma_{n-1} - (\gamma - 1 + \gamma_{n-1})(\gamma_{n-2} + 1 - \delta) + \sigma, \qquad (12)$$

$$P_n = -4p(\gamma_{n-2} + 1 - \delta). \qquad (13)$$

For left-hand side termination of the derived series at $n = 0$ we should have $R_0 = 0$, $a_{-1} = a_{-2} = 0$, hence

$$\gamma_0 = 1 - \gamma. \qquad (14)$$

Thus, the expansion is explicitly written as

$$u = \sum_{n=0}^{\infty} a_n B_z(1 - \gamma + n, 1 - \delta). \qquad (15)$$

Since the incomplete Beta function is defined only when the argument $1 - \gamma + n$ has a positive real part, the derived expansion applies for $\text{Re}(\gamma) < 1$.

The series is terminated when $P_{N+2} = 0$ and $Q_{N+1}a_N + P_{N+1}a_{N-1} = 0$ for some non-negative integer $N$ [$a_N$ then being the last nonzero coefficient of series (15)]. The first condition is fulfilled when

$$\gamma + \delta - 2 = N. \tag{16}$$

For each fixed $N = 0, 1, 2, \ldots$, the second restriction results in a polynomial equation of $(N+1)$th order for the accessory parameter $\sigma$, defining, in general, $N+1$ values of $\sigma$ for which the termination of the series is realized. Below are the explicit forms of the one- and two-term solutions.

$$N = 0 \implies \delta = 2 - \gamma, \quad \sigma = 4p(\gamma - 1): \tag{17}$$

$$u = B_z(1 - \gamma, 1 - \delta), \tag{18}$$

$$N = 1 \implies \delta = 3 - \gamma, \quad \sigma^2 + (1 + 12p - 8p\gamma)\sigma + 16p^2(2 - 3\gamma + \gamma^2) = 0: \tag{19}$$

$$u = B_z(1 - \gamma, 1 - \delta) + \frac{\sigma - 4p(\gamma - 1)}{\gamma - 1} B_z(2 - \gamma, 1 - \delta). \tag{20}$$

Note, however, that (16) implies that the sum of the parameters of each involved Beta function is a negative integer:

$$(1 - \gamma + n) + (1 - \delta) = n - N, \quad 0 \leq n \leq N,$$

so that according to (4) the resultant solution is expressed in terms of elementary functions. As already mentioned above, this observation is common for all the finite-sum solutions (see the discussion of this point below).

**2.** It is possible to construct a different expansion using the recurrence relation between the *derivatives* of the incomplete Beta functions,

$$z u'_n = u'_{n+1}, \tag{21}$$

Indeed, replacing $u'_n$ in Eq.(7) by $z u'_{n-1}$ and putting now $\sigma = 0$, we get

$$\sum_n a_n \left[ (4pz(z-1) + (\gamma - 1 + \gamma_n)(z-1) + (\delta - 1 + \delta_0)z) u'_{n-1} + 4\alpha p u_n \right] = 0. \tag{22}$$

Also here, we have to put $\delta_0 = 1 - \delta$ which leads to the following three-term recurrence relation

$$R_n a_n + Q_n a_{n-1} + P_n a_{n-2} = 0, \tag{23}$$

with

$$R_n = (\gamma - 1 + \gamma_n)\gamma_{n-2}, \tag{24}$$

$$Q_n = 4p\gamma_{n-2} - (\gamma - 1 + \gamma_{n-1})(\gamma_{n-3} + 1 - \delta), \tag{25}$$

$$P_n = -4p(\gamma_{n-3} + 1 - \delta) - 4p\alpha. \tag{26}$$

Now we conclude that for left-hand side termination of this series at $n = 0$ again we should have $\gamma_0 = 1 - \gamma$. If the series is right-hand side terminated then it necessarily holds that

$$\gamma + \delta - \alpha - 1 = N. \tag{27}$$

This time, the one- and two-term solutions with the corresponding additional equation for the parameters of the Heun equation are written as

$$N = 0 \implies \delta = 1 - \gamma + \alpha, \quad p\gamma = 0: \tag{28}$$

$$u = B_z(1 - \gamma, 1 - \delta), \tag{29}$$

$$N = 1 \implies \delta = 2 - \gamma + \alpha, \quad 4p\gamma(4p(\gamma - 1) - \alpha) = 0: \tag{30}$$

$$u = B_z(1 - \gamma, 1 - \delta) - 4pB_z(2 - \gamma, 1 - \delta). \tag{31}$$

Since $\gamma = 0$ is forbidden, for $N = 0$ the only choice is $p = 0$ so that this case is an unnecessary specification ($\delta = 1 - \gamma + \alpha$) of the trivial case $p = \sigma = 0$, which is solved in terms of the incomplete Beta function (29) for any $\delta$. However, starting from $N = 1$ the results are not so trivial.

**3.** A further possibility opens when trying an expansion in terms of specific ***combinations*** of the incomplete Beta functions. We will now demonstrate this constructing an expansion for the case $4p\alpha = \sigma$. We start from the following ansatz

$$u = \sum_n a_n(Au_n + u_{n+1}), \quad u_n = B_z(\gamma_0 + n, \delta_0) \tag{32}$$

with a constant $A$ and use recurrence relation (21) to transform the term proportional to $zu'_n$ (keeping unchanged the term $zu'_{n+1}$). As a result, we get instead of equation (7)

$$\sum_n a_n A\left[(4pz(z-1) + (\gamma - 1 + \gamma_n)(z-1))u'_n + (4\alpha pz - \sigma)u_n\right] + \\ \sum_n a_n\left[(4pz(z-1) + (\gamma - 1 + \gamma_{n+1})(z-1) + (\delta - 1 + \delta_0)(z + A))u'_{n+1} + (4\alpha pz - \sigma)u_{n+1}\right] = 0 \tag{33}$$

so that when $\sigma = 4\alpha p$ we may put $A = -1$ and then divide the equation by $z - 1$ thus arriving at the following equation

$$-\sum_n a_n\left[(4pz + (\gamma - 1 + \gamma_n))u'_n + \sigma u_n\right] + \\ \sum_n a_n\left[(4pz + (\gamma - 1 + \gamma_{n+1}) + (\delta - 1 + \delta_0))u'_{n+1} + \sigma u_{n+1}\right] = 0. \tag{34}$$

Now, if

$$\delta_0 + \delta = 0 \tag{35}$$

by putting $u'_{n+1} = zu'_n$ in the second sum this equation is rewritten as

$$\sum_n a_n \left[ (4pz + (\gamma - 1 + \gamma_n))(z-1)u'_n + \sigma(u_{n+1} - u_n) \right] = 0 . \qquad (36)$$

Hence, we can apply recurrence relations (8) and (9) to express all the terms in terms of functions $u_n$. As a result, we arrive at the following three-term recurrence relation

$$R_n a_n + Q_n a_{n-1} + P_n a_{n-2} = 0, \qquad (37)$$

$$R_n = (\gamma - 1 + \gamma_n)\gamma_{n-1}, \qquad (38)$$

$$Q_n = -(\gamma - 1 + \gamma_{n-1})(\gamma_{n-2} + \delta_0) + \sigma + 4p\gamma_{n-1}, \qquad (39)$$

$$P_n = -4p(\gamma_{n-2} + \delta_0) - \sigma . \qquad (40)$$

The condition for left-hand side termination of the constructed series at $n = 0$ is again $\gamma_0 = 1 - \gamma$. As regards the right-hand side termination of the series, it coincides with (27)

$$\gamma + \delta - \alpha - 1 = N . \qquad (41)$$

The solution corresponding to $N = 0$ involves two incomplete Beta functions:

$$N = 0 \Rightarrow \delta = 1 - \gamma + \alpha, \ p\delta = 0: \qquad (42)$$

$$u = B_z(1 - \gamma, -\delta) - B_z(2 - \gamma, -\delta), \qquad (43)$$

and the solutions for $N = 1$ is written as a combination of three incomplete Beta functions:

$$N = 1 \Rightarrow \delta = 2 - \gamma + \alpha, \ 4p\delta(4p\delta - 4p + \alpha) = 0: \qquad (44)$$

$$u = B_z(1 - \gamma, -\delta) - \left(1 - \frac{4p(\delta - 1)}{\gamma - 1}\right) B_z(2 - \gamma, -\delta) - \frac{4p(\delta - 1)}{\gamma - 1} B_z(3 - \gamma, -\delta) . \qquad (45)$$

**4.** Thus, the usage of combinations of the incomplete Beta functions has led to new results. However, of course, the above development is not the only possible one. We will now present a further example of application of the combinations of the incomplete Beta functions treating the case $\alpha = 0$. Let us again apply the form $u = \sum_n a_n (A u_n + u_{n+1})$ with $A = -1$. For $\alpha = 0$ equation (33) is rewritten as

$$-\sum_n a_n \left[ (4p z(z-1) + (\gamma - 1 + \gamma_n)(z-1))u'_n - \sigma u_n \right] +$$
$$+ \sum_n a_n \left[ (4p z(z-1) + (\gamma + \delta - 2 + \delta_0 + \gamma_{n+1})(z-1))u'_{n+1} - \sigma u_{n+1} \right] = 0. \qquad (46)$$

This leads to a ***four-term*** recurrence relation:

$$R_n a_n + Q_n a_{n-1} + P_n a_{n-2} + L_n a_{n-3} = 0, \qquad (47)$$

where

$$R_n = (\gamma - 1 + \gamma_n)\gamma_{n-1}, \qquad (48)$$

$$Q_n = 4p\gamma_{n-1} - (\gamma - 1 + \gamma_{n-1})(\gamma_{n-2} + \delta_0) - (\gamma + \delta - 2 + \delta_0 + \gamma_n)\gamma_{n-1} + \sigma, \tag{49}$$

$$P_n = -4p(\gamma_{n-2} + \delta_0) - 4p\gamma_{n-1} + (\gamma + \delta - 2 + \delta_0 + \gamma_{n-1})(\gamma_{n-2} + \delta_0) - \sigma, \tag{50}$$

$$L_n = 4p(\gamma_{n-2} + \delta_0). \tag{51}$$

For left-hand side termination of the derived series at $n=0$ (i.e., $R_0 = 0$, $a_{-1} = a_{-2} = a_{-3} = 0$) we again get that we should have $\gamma_0 = 1 - \gamma$. Thus, the final expansion is explicitly written as

$$u = \sum_{n=0}^{\infty} a_n [B_z(1 - \gamma + n, \delta_0) - B_z(2 - \gamma + n, \delta_0)]. \tag{52}$$

When this series is terminated then $L_{N+3} = 0$ for some non-negative integer $N$:

$$2 - \gamma + N + \delta_0 = 0. \tag{53}$$

For each fixed $N = 0, 1, 2, \ldots$, two more restrictions should necessarily be imposed on the parameters of the problem. These conditions are written as

$$Q_{N+1} a_N + P_{N+1} a_{N-1} + L_{N+1} a_{N-2} = 0 \tag{54}$$

and

$$P_{N+2} a_N + L_{N+2} a_{N-1} = 0. \tag{55}$$

This set of equations specifies two of the parameters of the initial Heun equation (say, $\delta$ and $\sigma$). In general, the system possesses $(N+1)(N+2)/2$ solutions (compare with corresponding one-term expansion (15) when the termination of the series is possible in $N+1$ cases). As can be checked, these solutions present exactly the same set of solutions as the one derived by expansion (15) for $N$ equal to $0, 1, 2, \ldots, N+1$ and $N+2$. Thus, the derived expansion (52) provides an alternative representation for the finite-sum incomplete Beta-function solutions obtained from expansion (15). This observation may be useful for some applications. Here are the solutions for $N = 0$ and $N = 1$ together with the restrictions imposed:

$$N = 0 \implies \delta_0 = -2 + \gamma, \; Q_1 = 0, \; P_2 = 0: \tag{56}$$

$$u = B_z(1 - \gamma, -2 + \gamma) - B_z(2 - \gamma, -2 + \gamma) = \frac{z^{1-\gamma}(1-z)^{-1+\gamma}}{1-\gamma}, \tag{57}$$

$$N = 1 \implies \delta_0 = -3 + \gamma, \; Q_2 Q_1 - R_1 P_2 = 0, \; P_3 Q_1 - R_1 L_3 = 0: \tag{58}$$

$$u = [B_z(1 - \gamma, -3 + \gamma) - B_z(2 - \gamma, -3 + \gamma)]$$
$$+ \frac{\sigma - (4p + 3 - \gamma - \delta)(\gamma - 1)}{\gamma - 1}[B_z(2 - \gamma, -3 + \gamma) - B_z(3 - \gamma, -3 + \gamma)]. \tag{59}$$

**5.** Several other developments are also possible. For instance, new possibilities open when one preliminarily transforms the initial Heun equation via change of the independent or dependent variables. Consider, for example, the transformation of the dependent variable

$$u = e^{sz} v(z) \tag{60}$$

resulting in the following equation

$$v'' + \left(2s + 4p + \frac{\gamma}{z} + \frac{\delta}{z-1}\right)v' + \frac{Az(z-1) + Bz + C}{z(z-1)} v = 0 \tag{61}$$

with

$$A = s(s + 4p), \quad B = 4p\alpha + s(\gamma + \delta), \quad C = -(s\gamma + \sigma). \tag{62}$$

The choice $s = -4p$ ($A = 0$) turns equation (61) into a new confluent Heun equation so that above developments apply here as well. However, of course, this is a rather trivial case. More elaborate cases come up when choosing $Az(z-1) + Bz + C \sim z^2$ or $z(z-1)$. Consider, for example, the first case, i.e., $B = A, C = 0$, achieved when

$$\sigma^2 + \gamma(-4p + \gamma + \delta)\sigma - 4p\alpha\gamma^2 = 0, \tag{63}$$

$$s = -\sigma/\gamma. \tag{64}$$

Trying now a Beta-function expansion for $v$ in the form of (2):

$$v = \sum_n a_n v_n, \quad v_n = B_z(\gamma_0 + n, \delta_0), \tag{65}$$

we will get

$$\sum_n a_n \left[ ((2s + 4p)z(z-1) + (\gamma - 1 + \gamma_n)(z-1) + (\delta - 1 + \delta_0)z)v_n' + Az^2 v_n \right] = 0. \tag{66}$$

Using further the recurrence relation between the derivatives of the incomplete Beta functions (21), we can replace $v_n'$ by $z^2 v_{n-2}'$ and then divide the equation by $z^2$, thus arriving at the situation already discussed in section 1. Note that Eq. (63) defines two new rather non-trivial values of $\sigma$ for which incomplete Beta-function expansions apply.

Finally, note that the other choice $Az(z-1) + Bz + C = Az(z-1)$, i.e., $\sigma = 4p\alpha\gamma/(\gamma + \delta)$ and $s = -\sigma/\gamma$, leads to an expansion in terms of combinations of incomplete Beta functions analogous to that considered above in section 3.

**6.** Let us now discuss if it is possible to construct finite-sum incomplete Beta-function solutions that are irreducible to elementary functions. The answer is no. Actually, this conclusion applies to a very large class of equations, not only to the confluent Heun equation discussed here.

A key observation to show this is that, in the force of the recurrence relation (8) [or (9)] and equation (21), any linear combination of *finite* number of incomplete Beta functions having the form $u_n = B_z(\gamma_0 + n, \delta_0)$, $n = 0,1,2,...$ can eventually be written as

$$u = \sum_{n=0}^{N} a_n u_n = A u_0 + \varphi(z) u_0', \qquad (67)$$

where $A$ is a constant, $\varphi(z)$ is an elementary function and $u_0 = B_z(\gamma_0, \delta_0)$; $u_0$ obeys the equation

$$u_0'' + \left(\frac{1-\gamma_0}{z} + \frac{1-\delta_0}{z-1}\right) u_0' = 0. \qquad (68)$$

Note that equation (67) shows that the finite-sum incomplete Beta-function solutions actually present a form of the Darboux transformation [11]. An important further observation following from equations (67) and (68) is that all the derivatives $u', u'', u''',....$ are proportional to $u_0'$ and thus are elementary functions since $u_0'$ is an elementary function, indeed,

$$u_0' = B_z'(\gamma_0, \delta_0) = z^{\gamma_0 - 1}(1-z)^{\delta_0 - 1}. \qquad (69)$$

If $A$ in equation (67) is zero, then $u$ is an elementary function. Suppose now that $A \neq 0$ and consider a *K*th order ordinary differential equation (not necessarily linear) of the following special form

$$F(z, u', u'', ...., u^{(K)}) + u = 0, \qquad (70)$$

where $F$ is an arbitrary elementary function of its arguments [evidently, the confluent Heun equation (1) is a particular simple case of this large class of equations]. In the force of what was said above about the derivatives of $u$, it follows that $F$ is an elementary function of $z$, whereby it is readily understood that $u$ is necessarily an elementary function.

Thus, a solution of equation (69) having the form of a linear combination of a *finite* number of incomplete Beta functions $B_z(\gamma_0 + n, \delta_0)$ is always reduced to elementary functions. Note that, in general, these are not polynomials. In our case of the confluent Heun equation these solutions present alternative representations of the known quasi-polynomial solutions [2]. As regards *infinite* series solutions presented here, these are new solutions not reported so far in the literature. We would like to note here that these solutions are of intrinsic importance and may be useful for different applications. A representative physical example is the application of these infinite series solutions to the optical surface polariton problem discussed in our earlier paper [9].

Consider, finally, the convergence of the derived infinite series. It follows from the above recurrence relations for the coefficients of the series that in all the presented cases holds

$$\lim_{n\to\infty} \frac{a_{n+1}}{a_n} = 1. \tag{71}$$

Noting now that for the incomplete Beta-functions $u_n = B_z(\gamma_0 + n, \delta_0)$

$$\lim_{n\to\infty}\left|\frac{u_{n+1}(z)}{u_n(z)}\right| = |z| \quad \text{and} \quad \lim_{n\to\infty}\left|\frac{u_{n+2}(z) - u_{n+1}(z)}{u_{n+1}(z) - u_n(z)}\right| = |z|, \tag{72}$$

it is easily seen (say, using the D'Alambert criterion) that all the derived series absolutely converge for $|z| < 1$. Further, using Gauss's test, it can be shown that all the series converge at $z = 1$ if $\text{Re}(\delta) < 1$ (note that the series apply for $\text{Re}(\gamma) < 1$).

**Acknowlegments**

I am grateful to Dr. David Melikdzanian for useful discussions. This work has been supported by the Armenian National Science and Education Fund (ANSEF Grant No. PS-10-2005) and RA (Grant No. 0074-2005).